\documentclass[aps,prl,twocolumn,superscriptaddress]{revtex4}

\usepackage[latin9]{inputenc}
\usepackage{amsmath}
\usepackage{amssymb}
\usepackage{graphicx}
\usepackage{makecell}
\usepackage{array}
\usepackage{CJK}

\makeatletter
\usepackage{color}%\usepackage{amstext}
\newcommand{\ehbar}{\hbar_{\mathrm{eff}}}
\usepackage{txfonts}

\makeatother

\begin{document}

\title{Quadratic growth of Out-of-time ordered correlators in quantum kicked rotor model}

\author{Guanling Li}
\affiliation{School of Science, Jiangxi University of Science and Technology, Ganzhou 341000, China}

\author{Wen-Lei Zhao}
\email[]{wlzhao@jxust.edu.cn}
\affiliation{School of Science, Jiangxi University of Science and Technology, Ganzhou 341000, China}

\begin{abstract}
We investigate both theoretically and numerically the dynamics of Out-of-Time-Ordered Correlators (OTOCs) in quantum resonance condition for a kicked rotor model. We employ various operators to construct OTOCs in order to thoroughly quantify their commutation relation at different time, therefore unveiling the process of quantum scrambling. With the help of quantum resonance condition, we have deduced the exact expressions of quantum states during both forward evolution and time reversal, which enables us to establish the laws governing OTOCs' time dependence. We find interestingly that the OTOCs of different types increase in a quadratic function of time, breaking the freezing of quantum scrambling induced by the dynamical localization under non-resonance condition. The underlying mechanism is discovered and the possible applications in quantum entanglement are discussed.
\end{abstract}
\date{\today}

\maketitle

\section{Introduction}

Quantum scrambling, a fundamental concept elucidating the spread of information across multiple degrees of freedom that is inaccessible via local measurements, has garnered extensive attention in quantum information~\cite{BYan20prl,BYan20prl2,JHWang22prr}, quantum chaos~\cite{Omanakuttan23,Sreeram21}, and condensed matter physics~\cite{SKZhao22prl,Dag19prl}. It is well known that the Out-of-time ordered correlators (OTOCs) can quantify the process of information scrambling with relevance to the operator growth~\cite{CYin21,PfZhang23prl}. The exponential growth of OTOCs, facilitated by exponential instability of chaos, produces the boundary of light cone of information scrambling in many-body systems~\cite{FLiu18prl,Das18prl}, for which the butterfly velocity of scrambling is closely related to quantum Lyapunov exponent~\cite{Keselman21prb,Mezei20JHEP,Pappalardi23entr}. The relaxation of OTOCs can detect the character of both the quantum thermalization and quantum entanglement~\cite{MacCormack21prb} in many-body systems, providing insights into the underlying connection between quantum chaos and information scrambling~\cite{Bilitewski21prb,TXu20prl}. Interestingly, genuine quantum chaos, specifically the superexponential instability induced by delta-kicking modulation in nonlinear interactions, can cause the superexponential growth of OTOCs~\cite{Wlzhao23arx1}, representing a new phenomenon of information scrambling~\cite{WLZhao21prb,ZQi23}.

The variants of the quantum kicked rotor (QKR) model under resonance conditions serve as ideal platforms to explore fascinating physics of quantum coherence~\cite{Santhanam22}, which has significant implications to the fundamental aspects of quantum transport~\cite{JWang09prl,Scoquart20prr} and topological new phases in Floquet systems~\cite{DHO12prl,Cheng19prl,LWZhou23arx,Longwen22prr}. The existence of flat band of quasi-energy spectrum determines the exponential diffusion dynamics in the on-resonance double-kicked rotor model~\cite{Hailong13pre}. The resonance condition yields Hofstadter's butterfly Floquet spectrum and topological phase transitions akin to the integer quantum Hall effect~\cite{Hailong13pre2,Raditya16pre,Longwen21Nano}, enriching our understanding of the quantum topological phenomena induced by chaos~\cite{YChen14prl}. Interestingly, the spinor QKR model with quantum resonance condition provides versatile playground to realize the quantum walk in momentum space~\cite{Gil16pra,Siamak18prl}, proposing a new protocol for the manipulation of the quantum transport with Floquet engineering~\cite{Marcel15pra}. The state-of-the-art experiments in atom-optics has indeed realized the QKR model and verified the dynamical phase transition and quantum walk therein by precisely tailoring the resonance condition for the driven period~\cite{Siamak19pra}. This paves the way for engineering exotic behavior of quantum information~\cite{Michele20jpb} and energy diffusion~\cite{Vakulchyk19prl} in various generalization of the QKR model.

In this context, we investigate both analytically and numerically the dynamics of different types of OTOCs under quantum resonance condition. The first type OTOCs $C_p$ involves two angular momentum operators. And the second one $C_T$ is constructed by the combination of the translation operator and angular momentum operator. We have derived the exact expression of the quantum state during both forward evolution and time reversal under quantum resonance conditions, which enables us to precisely establish the law governing the time dependence of OTOCs. Our findings reveal that both $C_p$ and $C_T$ exhibit unbounded quadratic growth, indicating a power law scrambling behavior in their long-term evolutions. The observation of similar time dependence laws for different OTOCs suggests a universality in this power law growth for the QKR model. It is known that the exotic physics exhibited by the QKR model under quantum resonance condition, such as ballistic energy diffusion and topologically-protected transport in momentum space, originates from the essential quantum coherence effects, without classical counterparts. Our findings unveil the role of quantum coherence in facilitating quantum scrambling, a connection of potential significance for applications in quantum information.

The paper is organized as follows. In Sec.~\ref{Sec-MResl} we describe the system and show the quadratic growth of OTOCs. In Sec.~\ref{TheoAnaly}, we show our theoretical analysis. A summary is presented in Sec.~\ref{Sum}.

\section{Model and main results}\label{Sec-MResl}
The dimensionless Hamiltonian of the QKR model reads
\begin{equation}\label{Hamila}
{\rm H}=\frac{{p}^2}{2}+ K\cos (\theta)\sum_n
\delta(t-t_n)\:,
\end{equation}
where $p=-i\ehbar\partial/\partial \theta$ is the angular momentum operator, $\theta$ is the angle coordinate, with commutation relation $[\theta,p]=i\ehbar$. Here, $\ehbar$ denotes the effective planck's constant, and $K$ is the kicking strength~\cite{Casati79}. One experimental realization of the QKR model involves ultracold atoms exposed to a pulsed laser standing field that mimics a delta-kicking potential~\cite{Moore1995}. The eigenequation of angular momentum operator is $p|\varphi_n\rangle = p_n |\varphi_n \rangle$ with eigenvalue $p_n = n\ehbar$ and eigenstate $\langle \theta|\varphi_n\rangle=e^{in\theta}/\sqrt{2\pi}$. With the complete basis of $|\varphi_n\rangle$, an arbitrary state can be expanded as $|\psi \rangle=\sum_n \psi_n |\varphi_n\rangle$. One period evolution of the quantum state from $t_n$ to $t_{n+1}$ is governed by $|\psi(t_{n+1})\rangle = U|\psi(t_{n+1})\rangle $. The Floquet operator $U$ involves two components, i.e., $U=U_fU_K$, where the $U_f =\exp\left(-ip^2/2\ehbar\right)$ represents the free evolution operator and the kicking term is denoted by $U_K =\exp\left[-i K\cos(\theta)/\ehbar\right]$.

The OTOCs are defined using the average of the squared commutator, i.e., $C(t)=-\langle[A(t),B]^2\rangle$. Here, both $A(t)=U^{\dagger}(t)A U(t)$ and $B$ are evaluated in Heisenberg picture. The average $\langle \cdot \rangle$ refers to the operator's expectation value concerning the initial state $\langle \psi(t_0)|\cdot|\psi(t_0)\rangle$~\cite{Li2017,Hashimoto2017,Mata2018,Zonnios22,WLZhao21prb}. We investigate two distinct OTOCs: one denoted as $C_p=-\langle[p(t),p]^2\rangle$, and the other as $C_{T}=-\langle[T\left(t\right),p]^2\rangle$, where $T=\exp\left({-i\epsilon p}/{\ehbar}\right)$ represents the translation operator. We focus solely on the quantum resonance condition, i.e., $\ehbar=4\pi$. Without loss of generality, we choose an initial state $\psi(\theta,t_0)=\cos(\theta)/\sqrt{\pi}$. Our main findings can be summarized by the following relationships
\begin{align}\label{cpotocequ}
\begin{aligned}
C_p(t)=12\pi^2K^2t^2\:,
\end{aligned}
\end{align}
and
\begin{align}\label{tpotocequ}
\begin{aligned}
C_{T}(t)=\sin^2\left(\frac{\epsilon}{2}\right)\left[2+\cos(\epsilon)\right]K^2t^2\:.
\end{aligned}
\end{align}
These relations clearly demonstrate the existence of the quadratic growth of different OTOCs.

\begin{figure}[t]
\begin{center}
\includegraphics[width=8cm]{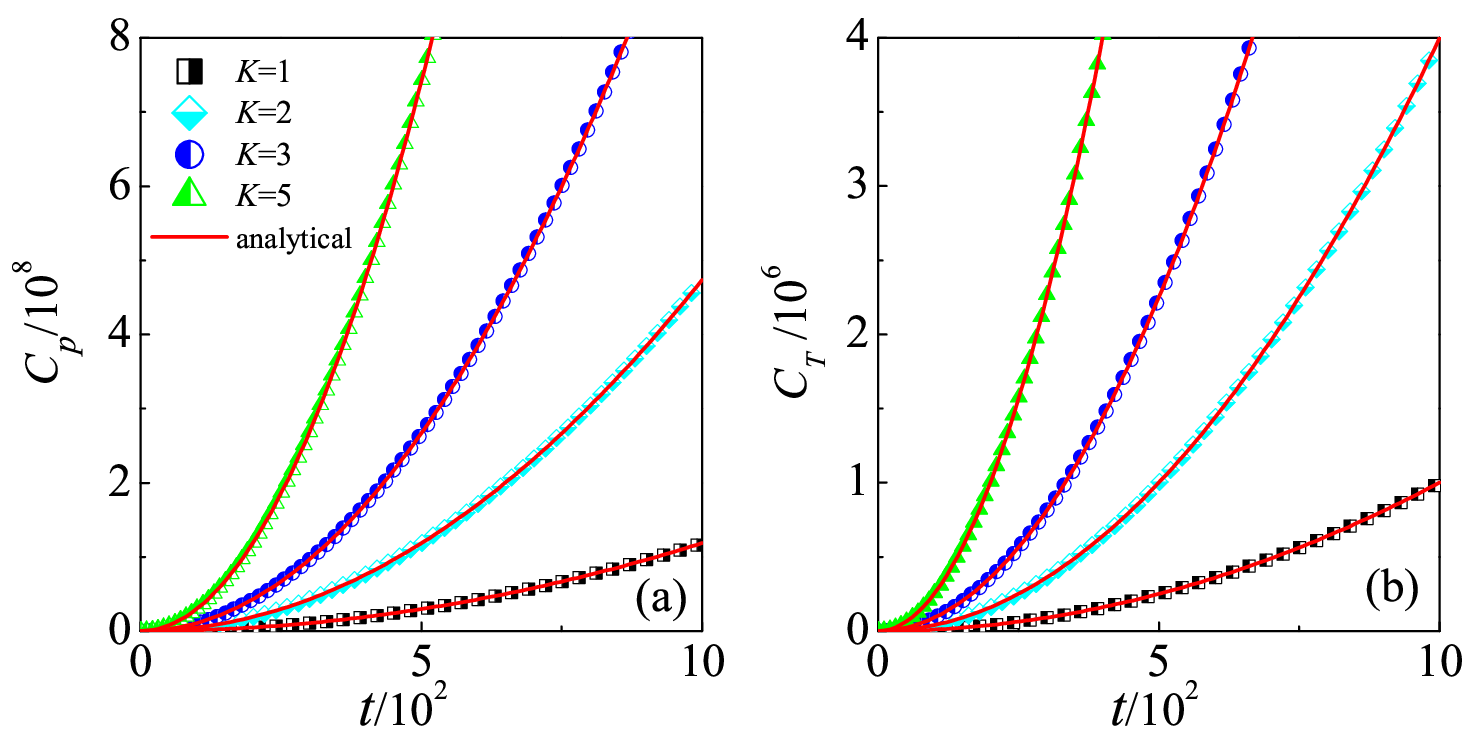}
\caption{Time dependence of the $C_p$ (a) and $C_{T}$ (b) for $K=1$ (squares), $2$ (diamonds), $3$ (circles), and $5$ (triangles). Red lines in (a) and (b) indicate our theoretical prediction in Eqs.~\eqref{cpotocequ} and ~\eqref{tpotocequ}, respectively. In (b), the value of the translation parameter is $\epsilon=\pi$.\label{OTOCST}}
\end{center}
\end{figure}

In order to confirm our above theoretical predictions, we numerically calculate both the $C_p$ and $C_T$ for different $K$. Our results demonstrate that for a specific $K$ (e.g., $K=1$ in Fig.\ref{OTOCST}), both $C_p$ and $C_T$ increase unboundedly with time. Furthermore, the larger the $K$, the faster they increase, following perfectly with the relations described in Eqs.\eqref{cpotocequ} and ~\eqref{tpotocequ}. It's noteworthy that the dependency of $C_T$ on the parameter $\epsilon$ offers a means of manipulating quantum scrambling by adjusting the translation operator, shedding light on the quantum control of non-Hermitian Floquet systems.
%%%%%%%%%%%%%%%%%%%%%%%%

The quadratic growth in OTOCs also emerges when we use the translation operator $T=\exp(-i\epsilon p/\ehbar)$ and a projection operator onto an initial state $B=|\psi(t_0)\rangle\langle\psi(t_0)|$ for OTOCs. In this situation, one can get the relation $C(t)=1-\mathcal{F}_O$, with $\mathcal{F}_O=|\langle \psi(t_0)|T|\psi(t_0)\rangle|^2$ being named as fidelity out-of-time ordered correlators (FOTOCs). Under the condition $\epsilon / \ehbar \ll 1$, straightforward derivation yields the approximation $C(t)\approx (\epsilon / \ehbar)^2 \left[\langle p^2(t)\rangle - \langle p(t)\rangle^2\right]$, by neglecting the terms in the Taylor expansion of $T=\exp(-i\epsilon p/\hbar)$ of orders larger than two. The mean momentum is zero, i.e., $\langle p(t)\rangle=0$ due to the symmetry of both the specific initial state $\psi(\theta,t_0)=\cos(\theta)/\sqrt{\pi}$ and the kicking potential. Therefore, the OTOCs is proportional to the mean energy, i.e., $C(t)\approx (\epsilon / \ehbar)^2 \langle p^2(t)\rangle\approx(\epsilon K t/2 \ehbar)^2$, indicating clearly the quadratic growth. Note that the Fourier spectrum of the FOTOCs $\mathcal{F}_O$ can be utilized in constructing the R{\'e}nyi entropy~\cite{Fan17SB,Lewis-Swan19nc,Garttner18prl}. In fact, FOTOCs have been used to characterize the multiple entanglements among different degrees of freedom in the kicked top model, which can be regarded as a collective of many spins~\cite{SCLi21pra}. It is known that the QKR model is a limit of the kicked top model with angular momentum being infinity~\cite{Haake88EL}. This provides the theoretical foundation for the significant implications of the quadratic growth of OTOCs in measuring the buildup of quantum entanglement.

\section{Theoretical analysis}\label{TheoAnaly}

It is straightforward to derive the relation
\begin{equation}\label{otoc}
C(t)= C_1(t)+ C_2(t)-2{\rm Re}\left[C_3(t)\right]\;,
\end{equation}
where the first two terms on the right side, i.e., two-points correlator, are defined as
\begin{align}\label{PartF}
C_{1}(t)& \mathrel{\mathop:}= \langle A^{\dagger}(t)B^2A(t) \rangle=\langle \psi_R(t_0) | B^2 |\psi_R(t_0)\rangle\;,
\end{align}
\begin{align}\label{SParta}
C_{2}(t) &\mathrel{\mathop:}= \langle B^{\dagger} A^{\dagger}(t)A(t)B\rangle =\langle \varphi_R(t_0)|\varphi_R(t_0)\rangle\;,
\end{align}
and the four-point correlator is given by
\begin{align}\label{TParta}
C_{3}(t) &\mathrel{\mathop:}= \langle A^{\dagger}(t)B A(t)B \rangle=\langle \psi_R(t_0)|B|\varphi_R(t_0)\rangle\;,
\end{align}
with $|\psi_R(t_0)\rangle = U^{\dagger}(t)AU(t)|\psi(t_0)\rangle$ and $|\varphi_R(t_0)\rangle =U^{\dagger}(t)AU(t)B|\psi(t_0)\rangle$. Here, $\rm {Re}\left[\cdots\right]$ denotes the real part of the complex variable~\cite{WLZhao21prb}.

The derivation of $C_1$ at a specific time $t=t_n$ involves three sequential steps~\cite{WLZhao23arx}. Firstly, evolving the initial state $|\psi(t_0)\rangle$ from $t_0$ to $t_n$ yields $|\psi(t_n)\rangle=U(t_n,t_0)|\psi(t_0)\rangle$. Secondly, applying operator $A$ to $|\psi(t_n)\rangle$ produces $|\tilde{\psi}(t_n)\rangle=A|\psi(t_n)\rangle$. Finally, the time reversal from $t_n$ to $t_0$ for $|\tilde{\psi}(t_n)\rangle$ results in $|\psi_R(t_0)\rangle=U^{\dagger}(t_n,t_0)|\tilde{\psi}(t_n)\rangle$. Equation~\eqref{PartF} indicates that $C_1$ is the expectation value of operator $B^2$ for $|\psi_R(t_0)\rangle$. The process to derive $C_2$ at time $t=t_n$ involves four steps. Firstly, applying the operator $A$ to the initial state $|\psi(t_0)\rangle$ yields the state $|\varphi(t_0)\rangle = A|\psi(t_0)\rangle$. Secondly, the forward evolution for the state $|\varphi (t_0)\rangle$ results in a state $|\varphi(t_n)\rangle=U(t_n,t_0)|\varphi(t_0)\rangle$. In the third step, we apply the operator $A$ to the state $|\varphi(t_n)\rangle$, which creates a new state $|\tilde{\varphi}(t_n)\rangle=A|\varphi(t_n)\rangle$. The fourth step involves time reversal for the state $|\tilde{\varphi}(t_n)\rangle$, giving $|\varphi_R(t_0)\rangle=U^{\dagger}(t_n,t_0)|\tilde{\varphi}(t_n)\rangle$. The norm of $|\varphi_R(t_0)\rangle$ defines $C_2$ as shown in Equation~\eqref{SParta}. With the two states $|\psi_R(t_0)\rangle$ and $|\varphi_R(t_0)\rangle$, we can calculate the $C_3$ based on Eq.~\eqref{TParta}.

Under the quantum resonance condition $\ehbar=4\pi$, each matrix element of  the free evolution operators $U_f$ in angular momentum space equal to unity, i.e., $U_f(n)=\exp(-i2\pi n^2)=1$. Consequently, these operators have no impact on the time evolution of quantum states. For one period evolution from $t=t_n$ to $t=t_{n+1}$, we only need to use the kicking evolution operator to act on the quantum state, i.e., $|\psi(t_{n+1})\rangle=U_K|\psi(t_{n})\rangle$. This leads to an exact expression of a quantum state at arbitrary time $t=t_n$ in angle coordinate space, i.e., $\psi(\theta,t_n) = U_K(\theta,t_n)\psi(\theta,t_0)=\exp[-i K t_n\cos(\theta)/\ehbar]\psi(\theta,t_0)$. Based on this, we can derive analytical expressions for both $|\psi_R(t_0)\rangle$ and $|\varphi_R(t_0)\rangle$, which yields the theoretical predictions for the OTOCs $C(t)$.

\subsection{Derivation of the $C_p$}

Given the operates $(A=p,B=p)$ and the quantum resonance condition, the three components of the OTOCs $C_p$ are denoted as $C_{p,1}(t)=\langle \psi_R(t_0) | p^2 |\psi_R(t_0)\rangle$, $C_{p,2}(t) =\langle \varphi_R(t_0)|\varphi_R(t_0)\rangle$, and $C_{p,3}(t)=\langle \psi_R(t_0)|p|\varphi_R(t_0)\rangle$, with $|\psi_R(t_0)\rangle = U_K^{\dagger}(t)pU_K(t)|\psi(t_0)\rangle$ and $|\varphi_R(t_0)\rangle =U_K^{\dagger}(t)pU_K(t)p|\psi(t_0)\rangle$.
At the time $t=t_n$, the action the operator $p$ to the state $\psi(\theta,t_n)=U_K(\theta,t_n)\psi(\theta,t_0)$ yields a new state $\tilde{\psi}(\theta,t_n)=p\psi(\theta,t_n)=\sin(\theta)\psi(\theta,t_n)Kt_{n}-i4\pi\psi^{(1)}\left(\theta, t_{0}\right)\exp\left[-i Kt_{n}\cos(\theta)/4\pi\right]$, where superscript $(n)$ ($n=1,2\ldots$) denotes the $n$-th order derivative of the functions. We then perform the time reversal from $t_n$ to $t_0$ starting from $\tilde{\psi}(\theta,t_n)$ and obtain
\begin{equation}\label{reQS-pt0-MT}
\begin{aligned}
\psi_R(\theta,t_0)&=\left[U_K(\theta,t_n)\right]^{\dagger}\tilde{\psi}(\theta,t_n)\\
&=Kt_{n}\sin(\theta)\psi(\theta,t_0)-i4\pi\psi^{(1)}\left(\theta, t_{0}\right)\:.
\end{aligned}
\end{equation}
With this state, one can get the analytical expression of $C_{p,1}(t_n)$
\begin{equation}\label{C1a}
\begin{aligned}
C_{p,1}(t_n)&=16\pi^2K^{2}t_n^{2}\int_{0}^{2\pi}|\Psi\left(\theta\right)|^2d\theta\\
&+256\pi^4\int_{0}^{2\pi}|\psi^{(2)}\left(\theta, t_{0}\right)|^2d\theta\;,
\end{aligned}
\end{equation}
where the function $\Psi(\theta)$ take the forms $\Psi\left(\theta\right)=\psi(\theta,t_0)\cos(\theta)+\psi^{(1)}\left(\theta, t_{0}\right)\sin(\theta)$.

For the derivation of $C_{p,2}(t_n)$, we apply the operator $p$ to acting on the initial state, which yields $\varphi(\theta,t_0)=p\psi(\theta,t_0)=-i4\pi\psi^{(1)}\left(\theta, t_{0}\right)$. Then, forward evolution from $t_0$ to $t_n$ creates the state $\varphi(\theta,t_n)=-i4\pi\psi^{(1)}\left(\theta, t_{0}\right)\exp\left[-i Kt_{n}\cos(\theta)/4\pi\right]$, along with $\tilde{\varphi}(\theta,t_n)=p\varphi(\theta,t_n)=Kt_{n}\sin(\theta)\varphi(\theta,t_n)-i4\pi\varphi^{(1)}\left(\theta, t_{0}\right)\exp\left[-\frac{i}{4\pi}Kt_{n}\cos(\theta)\right]$. Conducting the backward evolution from $t_n$ to $t_0$ for the state $\tilde{\varphi}(\theta,t_n)$, we obtain
\begin{equation}\label{repQSt0-c2-MT}
	\begin{aligned}
\varphi_R(\theta,t_0)=Kt_n\sin(\theta)\varphi(\theta,t_0)-i4\pi\varphi^{(1)}\left(\theta, t_{0}\right)\:.
	\end{aligned}
\end{equation}
With the assistance of the two states, we establish the following relations
\begin{equation}\label{c2a}
	\begin{aligned}
C_{p,2}(t_n)&=16\pi^2K^{2}t_n^{2}\int_{0}^{2\pi}|\psi^{(1)}\left(\theta, t_{0}\right)|^2\sin^2(\theta)d\theta+\\
&256\pi^4\int_{0}^{2\pi}|\psi^{(2)}\left(\theta, t_{0}\right)|^2d\theta\:,
	\end{aligned}
\end{equation}
and
\begin{equation}\label{real-C3a-0}
	\begin{aligned}
C_{p,3}(t_n)&=-16\pi^2K^{2}t_n^{2}\int_{0}^{2\pi}\Gamma\left(\theta\right)d\theta+ i64\pi^3 Kt_n \int_{0}^{2\pi} \Upsilon(\theta)d\theta\\
&-256\pi^4\int_{0}^{2\pi}\left[\psi^{(1)}\left(\theta, t_{0}\right)\right]^{\ast}\psi^{(3)}\left(\theta, t_{0}\right)d\theta\:.
	\end{aligned}
\end{equation}
Here, the superscript $*$ indicates the complex conjugate of the variable. The functions $\Gamma(\theta)$ and $\Upsilon(\theta)$ take the forms $\Gamma\left(\theta\right)=\psi^{\ast}\left(\theta,t_0\right)\left[\sin^2(\theta)\psi^{(2)}\left(\theta, t_{0}\right)+\frac{1}{2}\sin(2\theta)\psi^{(1)}\left(\theta, t_{0}\right)\right]$ and $\Upsilon(\theta)=\sin(\theta)\left[\psi^{\ast}\left(\theta,t_0\right)\psi^{(3)}\left(\theta, t_{0}\right)-\left[\psi^{(1)}\left(\theta, t_{0}\right)\right]^{\ast}\psi^{(2)}\left(\theta, t_{0}\right)\right]-\cos(\theta)|\psi^{(1)}\left(\theta, t_{0}\right)|^2$. Therefore, we can obtain the expression of the OTOCs
\begin{equation}\label{otoc-overal-NHppa}
\begin{aligned}
C_p(t_n)=&C_{p,1}(t_n)+C_{p,2}(t_n)-2\text{Re}[C_{p,3}(t_n)]\\
=&16\pi^2K^{2}t_n^{2}\int_{0}^{2\pi}\left\{\Phi\left(\theta\right)+2\text{Re}\left[\Gamma\left(\theta\right)\right]\right\}d\theta\\
&+128\pi^3Kt_n\int_{0}^{2\pi} \text{Im}\left[\Upsilon(\theta)\right]d\theta\\
&+512\pi^4\int_{0}^{2\pi}\text{Re}\left\{\left[\psi^{(1)}\left(\theta, t_{0}\right)\right]^{\ast}\psi^{(3)}\left(\theta, t_{0}\right)\right\}d\theta\\
&+512\pi^4\int_{0}^{2\pi}|\psi^{(2)}\left(\theta, t_{0}\right)|^2d\theta\:,
\end{aligned}
\end{equation}
with $\Phi\left(\theta\right)=\frac{1}{2}\left[|\psi^{(1)}\left(\theta, t_{0}\right)|^{2}\sin^2(\theta)+|\Psi\left(\theta\right)|^2\right]$ and $\text{Im}(\cdots)$ indicating the imaginary part of a complex variable. Substituting the initial state $\psi(\theta,t_0)=\cos(\theta)/\sqrt{\pi}$ into Eq.~\eqref{otoc-overal-NHppa} yields the equivalence
\begin{equation}\label{ppequa}
C_{p}(t)=12\pi^2K^2t^2\:.
\end{equation}

\subsection{Derivation of the $C_{T}$}

The three components of $C_T$ are represented as $C_{T,1}(t)=\langle \psi_R(t_0)|p^2|\psi_R(t_0)\rangle$, $C_{T,2}(t) =\langle \varphi_R(t_0)|\varphi_R(t_0)\rangle$, and $C_{T,3}(t)=\langle \psi_R(t_0)|p|\varphi_R(t_0)\rangle$. Here, the time-reversed states at time $t_0$, influenced by the operators $T=\exp(-i\epsilon p/\ehbar)$ and the initial states $\psi(t_0)$, take the forms $|\psi_R(t_0)\rangle = U_K^{\dagger}(t)\exp(-i\epsilon p/\ehbar)U_K(t)|\psi(t_0)\rangle$ and $|\varphi_R(t_0)\rangle =U_K^{\dagger}(t)\exp(-i\epsilon p/\ehbar)U_K(t)\exp(-i\epsilon p/\ehbar)|\psi(t_0)\rangle$, respectively. By repeating the same procedure for the derivation of both $|\psi_R(t_0)\rangle$ and $|\varphi_R(t_0)\rangle$ of $C_p$, we can obtain the exact expressions of the two states under quantum resonance condition
\begin{equation}\label{reQS-pt0tp-MT}
\begin{aligned}
\psi_R(\theta,t_0)=\psi\left(\theta+\epsilon,t_0\right)\exp\left[\frac{iKt}{2\pi}\sin\left(\frac{\epsilon}{2}\right)\sin\left(\frac{2\theta+\epsilon}{2}\right)\right]\:.
\end{aligned}
\end{equation}
and
\begin{equation}\label{repQSt0-c2tp-MT}
	\begin{aligned}
\varphi_R(\theta,t_0)=-i4\pi\psi^{(1)}\left(\theta+\epsilon, t_{0}\right)\exp\left[\frac{iKt}{2\pi}\sin\left(\frac{\epsilon}{2}\right)\sin\left(\frac{2\theta+\epsilon}{2}\right)\right]\:.
	\end{aligned}
\end{equation}
Consequently, one can derive analytically the three components of the $C_T$
\begin{equation}\label{C1tpa}
\begin{aligned}
C_{T,1}(t)&=4K^2t^2\sin^2\left(\frac{\epsilon}{2}\right)\int_{0}^{2\pi}\cos^2\left(\theta+\frac{\epsilon}{2}\right)|\psi\left(\theta+\epsilon, t_{0}\right)|^2d\theta\\
&+16\pi^2\int_{0}^{2\pi}\left|\psi^{(1)}\left(\theta+\epsilon, t_{0}\right)\right|^2d\theta\;,
\end{aligned}
\end{equation}
\begin{equation}\label{c2tpa}
	\begin{aligned}
C_{T,2}(t)&=16\pi^2\int_{0}^{2\pi} \left|\psi^{(1)}\left(\theta+\epsilon, t_{0}\right)\right|^2d\theta\:,
	\end{aligned}
\end{equation}
and
\begin{equation}\label{C3tpa}
	\begin{aligned}
C_{T,3}(t)=&i8\pi \sin\left(\frac{\epsilon}{2}\right)Kt\int_{0}^{2\pi}\upsilon\left(\theta\right)d\theta\\
&-16\pi^2\int_{0}^{2\pi}\psi^{\ast}\left(\theta+\epsilon, t_{0}\right)\psi^{(2)}\left(\theta+\epsilon, t_{0}\right)d\theta\:.
	\end{aligned}
\end{equation}
with $\upsilon\left(\theta\right)=\psi^{\ast}\left(\theta+\epsilon, t_{0}\right)\psi^{(1)}\left(\theta+\epsilon, t_{0}\right)\cos(\theta+\frac{\epsilon}{2})$. Combining these three parts yields
\begin{align}\label{otoc-overal-NHtpa-0}
\begin{aligned}
C_{T}(t)=&C_{T,1}(t)+C_{T,2}(t)-2\text{Re}[C_{T,3}(t)]\\
=&4K^2t^2\sin^2\left(\frac{\epsilon}{2}\right)\int_{0}^{2\pi}\cos^2(\theta+\frac{\epsilon}{2})|\psi\left(\theta+\epsilon, t_{0}\right)|^2d\theta\\
&-16\pi\sin\left(\frac{\epsilon}{2}\right)Kt\int_{0}^{2\pi}\text{Im}\left[\upsilon\left(\theta\right)\right]d\theta\;,\\
&+32\pi^2\int_{0}^{2\pi}\text{Re}\left[\psi^{\ast}\left(\theta+\epsilon, t_{0}\right)\psi^{(2)}\left(\theta+\epsilon, t_{0}\right)\right]d\theta\;,\\
&+32\pi^2\int_{0}^{2\pi}\left|\psi^{(1)}\left(\theta+\epsilon, t_{0}\right)\right|^2d\theta\;.
\end{aligned}
\end{align}
For a specific form of the initial state $\psi(\theta, t_{0})=\cos(\theta)/\sqrt{\pi}$, it is straightforward to establish the relation
\begin{align}\label{otoc-overal-NHtpa}
\begin{aligned}
C_{T}(t)=\sin^2\left(\frac{\epsilon}{2}\right)\left[2+\cos(\epsilon)\right]K^2t^2\:.
\end{aligned}
\end{align}

\section{Conclusion and discussions}\label{Sum}

In this work, we thoroughly investigate the dynamics of OTOCs, employing $C_p$ and $C_T$ under quantum resonance conditions.
The $C_p$ quantifies the commutation relation of two angular momentum operators at different times, while the $C_T$ measures that between the translation operator and angular momentum operator at different times. Our exact deductions of the quantum states during forward evolution and time reversal under quantum resonance allow us to establish the laws governing the time dependence of OTOCs. Our findings demonstrate that both $C_p$ and $C_T$ exhibit quadratic growth with time evolution, revealing an intrinsic power-law scrambling in their late-time behavior. Note that the mechanism of dynamical localization under non-resonant conditions suppresses quantum scrambling~\cite{Wlzhao23pra}. Therefore, the observed quadratic growth of OTOCs finds its origin in essential quantum coherence effects arising from quantum resonance, without classical analogs. We expect that the identification of similar power laws for different types of OTOCs reveals the universality in the power-law growth within the QKR model. Our discovery of the crucial role played by quantum coherence in facilitating quantum scrambling has significant implications in the fields of quantum information and quantum chaos.

\section*{ACKNOWLEDGMENTS}

This work is supported by the National Natural Science Foundation of China (Grant Nos. 12065009 and 12365002), the Natural Science Foundation of Jiangxi province (Grant Nos. 20224ACB201006 and 20224BAB201023).

\end{document}